%% file: main.tex
\documentclass[trackchanges, twocolumn]{aastex701}

\usepackage{amsmath}

\begin{document}



\title{Gravitational Wave Measurement of the \mmb\ Intrinsic Scatter at High Redshift}

\shorttitle{$M_\mathrm{BH}$--$M_\mathrm{bulge}$ Scatter Evolution}
\shortauthors{Matt et al.}

\correspondingauthor{Cayenne Matt}


\newcommand{\orcid}[1]{\href{https://orcid.org/#1}{\includegraphics[scale=0.04]{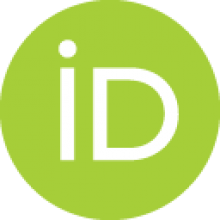}}}
\input{authors_scatter_arx}



\newcommand{\msigma}{$M_\mathrm{BH}$--$\sigma$}

\newcommand{\mmb}{$M_\mathrm{BH}$--$M_\mathrm{bulge}$}

\newcommand{\mone}{model M$_{\varepsilon}$}
\newcommand{\mtwo}{model M$_{\varepsilon +}$}
\newcommand{\mthree}{model M$_{\alpha}$}

\newcommand{\Mone}{Model M$_{\varepsilon}$}
\newcommand{\Mtwo}{Model M$_{\varepsilon +}$}
\newcommand{\Mthree}{Model M$_{\alpha}$}

\newcommand{\rev}{}
\newcommand{\revtwo}{}
\newcommand{\revthree}{}

\begin{abstract}

The observed GWB spectrum is higher in amplitude than model predictions by a factor of 2--3. Using a semi-analytic model, we evaluate the effect of a high-scatter supermassive black hole (SMBH) scaling relation (\mmb) on models of the \rev{nanohertz} gravitational wave background (GWB). By implementing an intrinsic scatter of the \mmb\ relation, which is larger at higher redshift, but matches local observations, we find that the amplitude of GWB models increases to be consistent with the low-frequency end of the GWB spectrum. This amplitude increase is not uniform across frequencies, a strongly evolving scatter preferentially increases the number density of the most massive SMBHs which, in the GWB spectrum, minimizes the strength of the low-frequency turnover. Our models with positively evolving intrinsic scatter can reproduce the electromagnetically observed overmassive SMBHs at $4 < z < 6$ without changing the \mmb\ normalization though we find that including moderate normalization evolution \revtwo{marginally} improves fits to the GWB data. We conclude that the \mmb\ relation which best describes the available GWB and electromagnetic data sets has intrinsic scatter that evolves as $\varepsilon(z) = \varepsilon_0 + (0.56 \pm 0.4) \log_{10}(1 + z)$ and normalization that evolves as $\alpha(z) = \alpha_0 (1 + z)^{0.84 \pm 0.35}$. The results of this work imply that the \mmb\ relation we see today is not universal throughout cosmic time and that a diversity of seeding models and growth mechanisms may be at play in the early stages of SMBH--galaxy evolution.


\end{abstract}



\section{Introduction} 

Accurate models for the supermassive black hole (SMBH) mass function \rev{outside the local universe} are important, but poorly defined. Locally, we see a strong correlation between SMBH mass and host galaxy bulge mass \citep{Kormendy_Ho_2013, McConnell_2013}. Unfortunately, electromagnetic (EM)-based SMBH mass measurements become prohibitively difficult for more distant galaxies meaning that correlations we observe locally, such as \mmb, are not well defined for higher redshifts. This relationship between SMBH and host bulge mass is a critical indicator of how galaxies and black holes evolve (together and/or separately). Where populations of SMBH--galaxy pairs lie on this relation through time provides insights into their rates of growth, feedback between growth mechanisms, and the relative importance and contributions of, e.g., accretion, star formation, and mergers for these objects.

Many studies have leveraged theory, EM observations, cosmological simulations, or gravitational waves to constrain the high-redshift location of this relation without general consensus \citep[e.g.][]{Wyithe_2003, Zhang_2023, Zou_2024, Chen_2024, Matt_2026}. A notable portion of these studies focus primarily on the normalization (y-intercept) of the \mmb\ relation. Some EM-based studies find evidence for an increased \mmb\ normalization in the past, which would imply a black-hole-first growth pathway\rev{, i.e., black hole ``dominance'', where the black hole growth initially outpaces that of the galaxy and/or black holes originate from heavy seeds \citep{Volonteri_2012, Zhuang_2023}}. Other studies find no evidence for changes to the normalization with time \citep{Mountrichas_2023, Zou_2024, Li_2025_normal, Sun_2025_ne, Geris_2026}. \rev{Since the launch of the \textit{James Webb Space Telescope} (JWST) in 2021, numerous studies have reported SMBH and galaxy mass measurements at $z > 4$ \citep[e.g.,][]{Ding_2023, Harikane_2023, Maiolino_2024, Li_2025_normal}, with no clear consensus on the intrinsic \mmb.} It is possible that the discrepancy between results could arise due to observational biases in some magnitude limited surveys which would leave lower mass SMBHs unobserved and thus artificially inflate the amplitude and flattening the relation \citep{Lauer_2007, Zhang_2023_Trinity, Li_2025_bias}. It has been suggested that there are lower mass SMBHs at high redshift that evade detection \citep{Shankar_2016, Shankar_2019, Silverman_2025, Ziparo_2026}.

Recently \citet{Li_2025_normal} selected AGN \rev{from archival JWST data} based on the presence of a broad H$\alpha$ component rather than using a magnitude / luminosity cutoff. The result of this selection was a population of SMBH--galaxy pairs with mass ratios consistent with the local AGN relations. This result suggests that many surveys that find exclusively overmassive SMBHs may be incomplete and the offset of the \mmb\ relation at \rev{$z > 4$} relative to today is less extreme (and possibly zero). If this is the case then perhaps the normalization is not higher\rev{;} however, the local relation still does not predict these extremely high mass SMBHs seen at \rev{$z > 4$}. An alternative explanation is that the intrinsic scatter of the \mmb\ relation was higher in the past therefore reproducing these overmassive black holes without needing extreme SMBH growth on a population level. This hypothesis simultaneously predicts a large population of normal and low-mass-ratio SMBH--galaxy pairs \citep{Brooks_2025, Li_2025_normal, Geris_2026}.

Locally, the \mmb\ relationship has relatively low intrinsic scatter \citep[second only to the \msigma\ relation;][]{Kormendy_Ho_2013, McConnell_2013, Ferrarese_2000, Gebhardt_2000, Gultekin_2009, Gultekin_2011, Tremaine_2002, Greene_2016}. This correlation implies a coupled growth pathway for SMBH--galaxy pairs and the scatter about this relationship reflects small differences in growth rates \citep{Zhuang_2023, Terrazas_2025, Izquierdo-Villalba_2026} and, at the low-mass end, seed information \citep[T. K. Waters et al. in prep; ][]{Greene_2020}. If the growth rates of both galaxies and their central SMBHs were to be roughly similar, then objects would grow along the \mmb\ relation with cosmic time. Galaxy--SMBH pairs would be born in the low-mass region of the relation and evolve up, following the slope of the relation until they reach their present-day masses. In this case, the \mmb\ relation would be a valuable tool for predicting SMBH masses based on galaxy bulge masses for distant galaxies where the SMBH is obscured and/or spectral resolution is too low to make direct SMBH mass measurements. Indeed some studies have found that both the normalization and scatter of the \mmb\ relation is unchanging with time \citep{Guia_2024, Zou_2024}, however these results fail to predict the massive SMBHs at \rev{$z > 4$}.

An increased intrinsic scatter could indicate that there is a diversity of growth pathways for SMBHs and galaxies \citep{Zhuang_2023, Hu_2025, Terrazas_2025, Izquierdo-Villalba_2026}. For example, overmassive SMBH hosts have high star formation rates so the galaxy can catch up in mass. Conversely undermassive SMBHs have low star formation rates and high Eddington ratios (and high AGN luminosities) because they are growing their SMBHs to catch up. Then, pairs can stay on the relation when they have positive star formation--accretion feedback and merger-based growth. Independently, an increased scatter \rev{in the past}, but low scatter locally is predicted even if there is no physical link between SMBH and host galaxy mass \citep{Jahnke_2011}. Furthermore, \citet{Hirschmann_2010} modeled the change in scatter because of mergers and found that scatter decreases with the number of mergers because of central limit theorem considerations. More recently \citet{Zhu_2025} compared the sources of intrinsic scatter through time in various cosmological simulations. Their results showed that the dominant source and scale of scatter of the \mmb\ relation can change both with redshift and mass indicating that the value of scatter outside the local universe can provide important insights into SMBH--galaxy coevolution. Other studies have found that some cosmological simulations can reproduce the overmassive population of SMBHs seen by \rev{JWST} with large values of intrinsic scatter \citep{Dattathri_2025}.

EM-based constraints on \revtwo{the \mmb\ relation} remain difficult. \rev{There are two major limitations that lead to under-detection of low $M_\mathrm{BH}/M_\star$ ratios: (i) Low luminosity AGN in massive galaxies are difficult to see because they are outshined by the host and \citep[e.g., ][]{Silverman_2025, Li_2025_bias, Volonteri_2023} and (ii) low luminosity AGN are difficult / impossible to find in magnitude-limited surveys \citep[e.g.,][]{Lauer_2007, Li_2025_bias, Schulze_2014}}. Gravitational waves produced by SMBH binaries do not face these same challenges. It has been long since theorized that we can get constraints from the \rev{nanohertz} gravitational wave background (GWB) \citep{Simon_2016}. There is now strong evidence of a detection of the GWB from pulsar timing arrays (PTA) \citep{Agazie_2023, Antoniadis_2023, Reardon_2023, Xu_2023}. Since the $4\sigma$ evidence announcement there has been a surge of analyses placing constraints on SMBH populations \citep{AgazieBHB_2023,  Gardiner_2023, Chen_2024, Sato-Polito_2024, Liepold_2024, Bonoli_2025, Matt_2026}. Though the details vary between these studies, the broad consensus is that the unexpectedly high GWB amplitude requires more high-mass SMBHs than predicted by the local \mmb\ relation. Some studies have focused on the SMBH--galaxy relations themselves, while others have revised our predictions for galaxy populations \citep{Sato-Polito_2024, Liepold_2024, Matt_2026}. There is now strong evidence to suggest that there are high-redshift SMBH--galaxy pairs which lie both above and on the local relation leading to conflicting narratives of how the \mmb\ relation may evolve \citep{Pacucci_2023, Li_2025_normal}. \rev{An \mmb\ relation with an intrinsic scatter that was higher at earlier epochs increases the number of high- and low-mass SMBHs relative to the local relation for $z > 0$.} This boost in high-mass SMBHs can amplify the predicted amplitude of the GWB and produce a population of overmassive SMBHs at high redshift while maintaining the same fundamental SMBH--galaxy mass relationship observed locally.

In this work we consider a model for the GWB which uses an \mmb\ relation with an evolving intrinsic scatter. We evaluate the potential of this model to unify EM and gravitational wave observations across a wide range of redshifts.

In \rev{Section} \ref{sec:sam} we describe our model setup and the functional form of scaling relations and parameter evolution. We describe the behavior of and effect of evolutionary parameters on our models in \rev{Section} \ref{sec:impact_of_epz}. The GWB-based results of our model fits are presented in \rev{Section} \ref{sec:results} while comparisons with EM data are presented in\rev{Section} \ref{sec:results_em_comp}. We place our findings into context in \rev{Section} \ref{sec:discussion} and a summary of this work can be found in \rev{Section} \ref{sec:summary}.

We assume WMAP9 cosmology \citep{wmap9} and all values and parameters reported in log space are base 10 unless otherwise specified.

\section{Semi-Analytic Models}\label{sec:sam}

We use a semi-analytic SMBH binary population synthesis code, \textsc{holodeck}, to calculate the expected GWB spectrum. We briefly describe the model \rev{setup} here though a description of the model workflow and, e.g., SMBH binary hardening models, can be found in \citet{AgazieBHB_2023} with updates and recent changes further detailed in \citet{holodeck} and \citet{Matt_2026}.

We assume an initial distribution of SMBH total mass ($M_{\mathrm{tot}})$, mass ratio ($q = m_1/m_2$), and redshift from which we define a population of merging galaxies based on the merger rate prescription in \citet{Rodriguez_Gomez_2015} and the galaxy stellar mass function (GSMF) in \citet{Leja_2020}. \rev{We then convert this distribution of galaxies into a population of SMBH binaries by assuming binaries are formed following a galaxy merger. Their masses are assigned based on the galaxy bulge masses via the \mmb\ relation. This gives the number density of SMBH binaries per unit mass, mass ratio, and redshift, which are then evolved from $\sim 10$ kpc separations down to the gravitational wave regime according to the binary hardening prescription in \citet{Kelley_2017, Agazie_2023}. The non-gravitational wave binary hardening processes include dynamical friction and stellar scattering, and we allow for hardening timescales long enough that some binaries effectively stall and not every galaxy merger results in a SMBH binary that contributes to the GWB. The GWB amplitude is then calculated following \citet{Phinney_2001}.}

Locally, the relationship between SMBH mass and galaxy bulge mass follows a power-law relation \citep[see, e.g.,][]{Kormendy_2001, Kormendy_Ho_2013, McConnell_2013}. In our model, we define the \mmb\ relation to be
\begin{equation}
    M_\mathrm{BH}=\alpha_0 \left(\frac{M_{\mathrm {bulge}}}{10^{11}\ M_{\odot}}\right)^{\beta_0}
	\label{eqn:mm}
\end{equation}
with an intrinsic scatter in $\log M_\mathrm{BH}$ drawn from a normal distribution \citep{Gultekin_2009} with standard deviation $\varepsilon_0$. In this paper we consider a scatter that is a function of redshift given by the form
\begin{equation}
    \varepsilon(z) = \varepsilon_0 + \varepsilon_z \log(1 + z),
	\label{eqn:epsilonz}
\end{equation}

where the $z = 0$ values can be determined from observation \citep[$\log \alpha_0 = 8.69$; $\beta_0 = 1.17$; $\varepsilon_0 = 0.28$ dex;][]{Kormendy_Ho_2013}, and $\varepsilon_z$ can be positive or negative with $\varepsilon_z = 0$ indicating no evolution in the relation. The intrinsic scatter in the relation is scatter in the SMBH masses; it is explicitly incorporated into the \mmb\ relation in \textsc{holodeck} and was a free parameter in both \citet{AgazieBHB_2023} and \citet{Matt_2026} though it was assumed to be unchanging with redshift. It has been suggested that the scatter in the \mmb\ relation may be different for high versus low mass SMBHs \citep{Greene_2020}. The available data from the GWB cannot currently constrain SMBH masses at $M_\mathrm{BH} \lesssim 10^{8} M_\odot$ \citep{AgazieBHB_2023}. Therefore we are only interested in the most massive SMBHs and a mass-independent intrinsic scatter is sufficient for our analysis.

\begin{deluxetable}{lllll}
\tablewidth{0pt}
\tablehead{
\colhead{Parameter} & \colhead{Purpose} & \colhead{\Mone} & \colhead{\Mtwo} & \colhead{\Mthree}
}
\startdata
\multicolumn{5}{c}{\mmb} \\
\hline
$\varepsilon_z$  & \rev{Scatter evolution} & $\mathcal{U}[-0.2, 2.0]$ & $\mathcal{U}[-0.2, 2.0]$ & 0.5 \\
$\alpha_z$       & \rev{Normalization evolution} & 0.0 & 0.0 & $\mathcal{U}[-3.0, 3.0]$ \\
$\log\alpha_0$   & \rev{Local normalization} & 8.69 & $\mathcal{N}[8.69, 0.05]$ & 8.69 \\
$\beta_0$        & \rev{Local slope} & 1.17 & $\mathcal{N}[1.17, 0.08]$ & 1.17 \\
\hline
\multicolumn{5}{c}{SMBH Binary Hardening} \\
\hline
$\tau_f$ [Gyr]   & \rev{Total binary lifetime} & 3.0 & $\mathcal{U}[0.1, 11.0]$ & 3.0 \\
$\nu_\mathrm{inner}$ & \rev{Stellar scattering exponent} & -1.0 & $\mathcal{U}[-2.0, 0.0]$ & -1.0 \\
$R_\mathrm{char}$ [pc]\tablenotemark{a} & \rev{Characteristic radius} & 10.0 & $\mathcal{U}[2.0, 20.0]$ & 10.0 \\
\hline
\multicolumn{5}{c}{Galaxy Stellar Mass Function} \\
\hline
$\phi_{*, 0, 0}$     & \rev{Local normalization of first Schechter} & -2.383 & $\mathcal{N}[-2.383, 0.28]$ & -2.383 \\
$\phi_{*, 1, 0}$     & \rev{Local normalization of second Schechter} & -2.818 & $\mathcal{N}[-2.818, 0.050]$ & -2.818 \\
$M_{\mathrm{c}, 0}$  & \rev{Local characteristic mass} & 10.767 & $\mathcal{N}[10.767, 0.026]$ & 10.767 \\
\enddata
\tablenotetext{a}{\citet{AgazieBHB_2023} assumed a fixed value of $R_\mathrm{char} = 100$ pc. In the absence of other changes, our lower value increases the strength of the low-frequency turnover of the GWB spectrum.}
\caption{A summary of the parameter \rev{setup} for the three models in this paper. We note fixed parameters by single values and free parameters by their prior distributions. Both \mone\ and \mthree\ had one free parameter while \mtwo\ allowed 9 parameters to vary. We used uniform distributions for the \mmb\ evolutionary parameters ($\varepsilon_z$ and $\alpha_z$) and SMBH binary hardening parameters. We use the same priors for $\tau_f$ and $\nu_\mathrm{inner}$ as in \citet{AgazieBHB_2023}. For the local \mmb\ and galaxy stellar mass function parameters we used \rev{Gaussian} priors based on \citet{Kormendy_Ho_2013} and \citet{Leja_2020}. All additional parameters not in this table are fixed to their fiducial value \citep{Matt_2026} for all three models.}
\label{tab:models}
\end{deluxetable}

\rev{For this work we consider three models, two with a variable scatter evolution power-law, $\varepsilon_z$, and one where $\varepsilon_z$ is fixed to 0.5. A summary of the model setups can be found in Table \ref{tab:models}. Our prior on $\varepsilon_z$ is uniform with $-0.2 \leq \varepsilon_z \leq 2.0$, as the $\varepsilon_z$ parameter is only physically meaningful inside this range. If $\varepsilon_z < -0.2$ the intrinsic scatter drops below $\varepsilon_z \sim 0.1$ by $z \sim 2$. With $\varepsilon_z > 2.0$ SMBHs get scattered outside the \rev{nanohertz} gravitational wave regime and roughly 30--40\% of galaxies with $M_\star > 10^{10} M_\odot$ have a SMBH--galaxy \rev{bulge} mass ratio above 1 by $z \sim 3$. There is sufficient evidence from EM observations to rule out a high redshift \mmb\ relation with lower intrinsic scatter than the local value of 0.28 dex, but we still allow our lower range to include some negative evolution for model freedom and so we can be sure that the result is purely driven by our data.}

\rev{In \mtwo\ we chose to sample three hardening parameters: the total hardening time, $\tau_f$, the stellar scattering \rev{power-law} slope, $\nu_\mathrm{inner}$, and the characteristic hardening radius $R_\mathrm{char}$. These parameters can have strong impacts on the shape of the GWB spectrum \citep{Kocsis_2011, Sesana_2013, AgazieBHB_2023, Laal_2025_priors}. As discussed in the following section, \mmb\ scatter evolution also has an impact on the GWB spectral shape. We therefore allowed these hardening parameters to vary for one model to investigate any possible degeneracies. We additionally sample three GSMF parameters: the two local normalizations $\phi_{*, 0, 0}$ and $\phi_{*, 1, 0}$ and the local characteristic mass $M_{\mathrm{c}, 0}$. Previous GWB studies have found degeneracies between GSMF and \mmb\ parameters \citep{AgazieBHB_2023, Matt_2026}.  To address these degeneracies, we fix them for \mone\ and \mthree, but we leave them variable in \mtwo\ to investigate how this affects our results.}

\rev{To determine our best-fit models and posteriors, we draw 20,000 samples \revtwo{directly} from our prior space \revtwo{using Latin hypercube sampling} and calculate the GWB spectrum for each sample instance using} \textsc{holodeck}. \rev{We determine goodness of fit by comparing each of the 20,000 spectra to the data and calculating a likelihood. The spectrum with the highest likelihood is our best-fit model. The likelihoods we calculate represent how well models fit the GWB data, but are not marginalized over the priors. Our posteriors are then the \revtwo{priors weighted by the likelihoods}. This method was used by \citet{Matt_2026} and was first described in the appendix of \citet{AgazieBHB_2023}.}

\section{Impact of Evolutionary Parameters}\label{sec:impact_of_epz}

\rev{To aid in the interpretation of results, we describe the isolated impact of two types of \mmb\ evolution on the GWB spectrum and black hole mass function (BHMF) in this section.} The GWB amplitude is sensitive to the number of the most massive SMBHs \citep{Phinney_2001} so changes to the \mmb\ relation that can increase the number of massive SMBHs will increase the predicted GWB amplitude in absence of any other change. Both an increased \mmb\ scatter and normalization will produce a greater number of massive SMBHs and so we would expect that either change could increase the predicted GWB amplitude.

\begin{figure*}[ht]\centering
    \includegraphics[width=\textwidth, keepaspectratio]{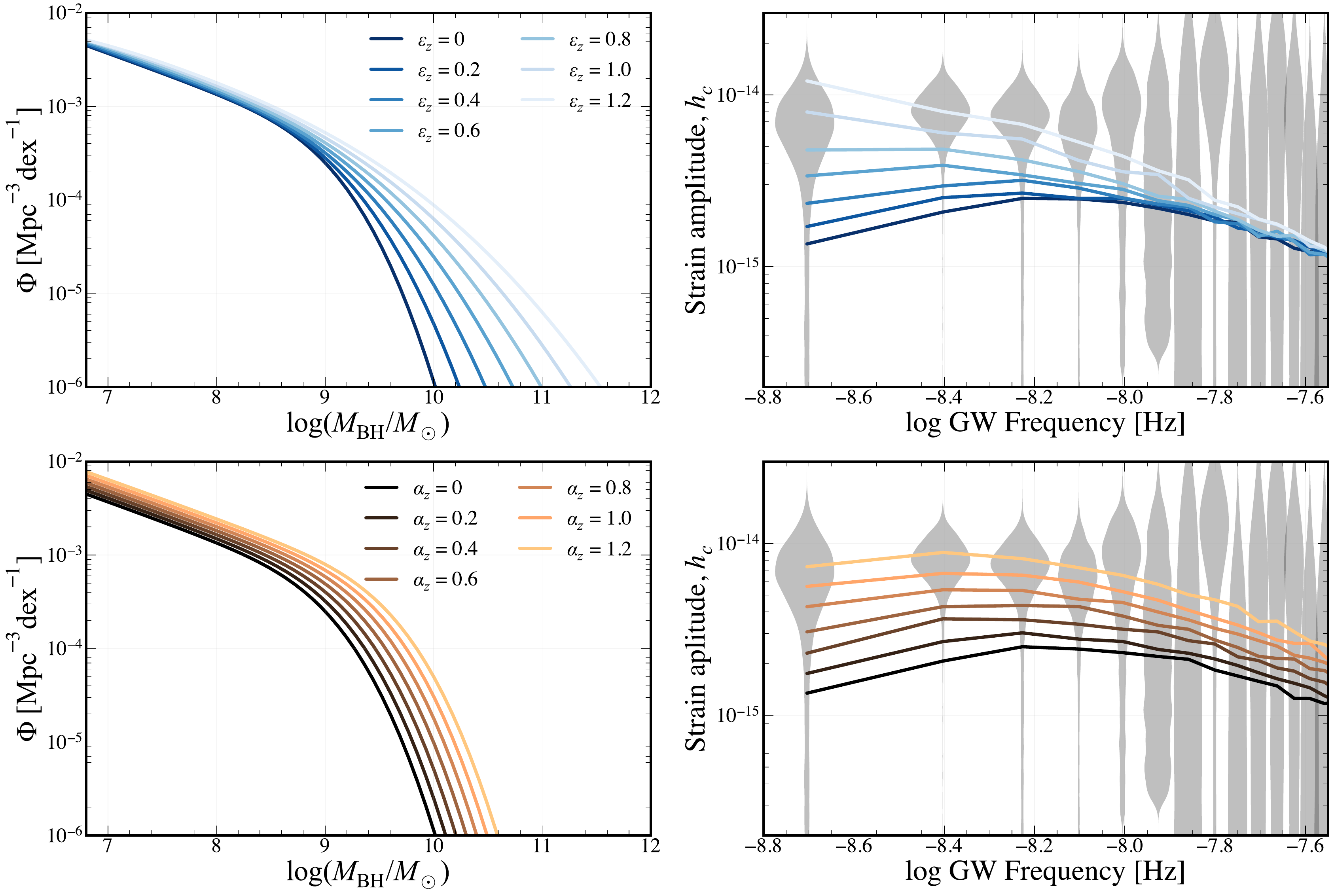}
    \caption{Here we show the separate effects of an evolving \mmb\ scatter (top) and normalization (bottom) on the $z = 2$ BHMF (left) and the GWB spectrum (right). We keep all parameters in the models constant and change only the power law for evolving scatter, $\varepsilon_z$ (or that for evolving normalization, $\alpha_z$) where $\varepsilon_z, \alpha_z = 0$ indicates no redshift evolution and $\varepsilon_z, \alpha_z > 0$ indicates positive redshift evolution. Both types of evolution increase the number density of the most massive SMBHs where the amount of the increase is correlated with the strength of evolution, but only an evolving normalization significantly increases the number density of SMBH masses $M_\mathrm{BH} < 10^{9} M_\odot$. Gravitational wave frequency is inversely proportional to SMBH chirp mass. Therefore the relative changes to number densities across the mass range have corresponding impacts on the GWB spectrum. Both types of evolution increase the amplitude at the lowest frequencies (high masses), but only $\alpha_z$ affects the high frequency (low mass) end of the spectrum.}
    \label{fig:epsilon_z_effects}
\end{figure*}

Intuitively, one might initially expect the impact of increasing scatter to be similar to increasing normalization, however we find the effects to be more complex. In Figure \ref{fig:epsilon_z_effects} we compare the effect of an evolving scatter (top row) with that of an evolving \mmb\ normalization (bottom row). In the left-hand column, we show the $z = 2$ BHMF as calculated using different values of $\varepsilon_z$ and $\alpha_z$ indicating different strengths of evolution in scatter and normalization. Here $\alpha_z$ is the normalization analog of $\varepsilon_z$ \citep[$\alpha(z) = \alpha_0 (1 + z)^{\alpha_z}$, i.e. the same evolutionary form as][]{Matt_2026}. The black / darkest blue line in every panel is the model with no redshift evolution of the \mmb\ relation. The lighter colors correspond to greater values of the corresponding evolutionary parameter and therefore represent \mmb\ relations with greater scatter / normalization at \rev{$z = 2$} than observed locally. \rev{We note that our fiducial (non-evolving) BHMF over predicts the SMBH number density compared to AGN-inferred BHMFs, a  discrepancy discussed in some detail by, e.g., \citet{Liepold_2024}.}

In the left-hand column of Figure \ref{fig:epsilon_z_effects} we see that both types of positive evolution increase the number density of the most massive SMBHs and the amount of the increase is correlated with the strength of evolution. An increased \mmb\ normalization is equivalent to increasing every calculated SMBH mass by some amount. \rev{The effect of increasing SMBH mass is to shift the entire BHMF to the right on the x-axis (lower-left panel)}. The horizontal difference between any two curves is roughly equivalent across the entire mass range, but the vertical difference (number density) is greater at high masses because of the steeper slope relative to the low-mass end. Increasing the intrinsic scatter shifts the SMBH masses both higher and lower so the relative shift along the x-axis is identical across the mass range. In fact, the greatest increase is to the highest mass number densities and there are only negligible changes to the low mass number densities. Though more commonly discussed in the context of observational bias, this effect the same as that which causes Eddington bias \citep{Eddington_1913}.

These parameters have differing effects at high and low masses which affects not only the predicted BHMF, but also the associated GWB spectrum. \rev{In the right-hand column of Figure \ref{fig:epsilon_z_effects} we show the GWB spectrum associated with the same values for evolutionary parameters as in the left-hand panel. Color coding is the same across columns. The GWB spectra are calculated with} \textsc{holodeck}\rev{ across $0 \leq z \leq 10$. For an equal mass circular binary, the frequency of gravitational waves emitted is twice the orbital frequency.} The orbital frequency of the binary is related to the masses of the SMBHs via Kepler's third law, so the lower frequencies of the GWB correspond to higher SMBH chirp masses \citep[$f_\mathrm{GW, obs} \propto (M (1 + z))^{-1}$, see][]{Phinney_2001}. Therefore the relative changes to number densities across the mass range have corresponding impacts on the GWB spectrum. Both types of evolution increase the GWB amplitude at the lowest frequencies (high masses), but only $\alpha_z$ affects the high frequency (low mass) end of the spectrum. The difference in GWB amplitudes is strongest for $\log f_\mathrm{GW} \gtrsim -8.0$ where the the GWB data are poorly constrained, but there are still notable differences across the range we fit to (lowest five frequency bins).

The GWB spectrum produced by a population of SMBH binaries which harden only via gravitational wave emission is a power law \citep{Phinney_2001}. Deviations from this power law can occur for a variety of reasons, but a turnover at the low-frequency end is expected if additional hardening mechanisms are present \citep{Kocsis_2011, Sesana_2013}. Previous studies have found that the GWB data are well described by a spectrum with a turnover due to environmental effects \citep{AgazieBHB_2023, Harris_2026}. The underlying model we used to produce this figure includes both stellar scattering and dynamical friction as hardening mechanisms prior to the gravitational wave emission phase. The turnover is apparent in the black (fiducial) spectrum by the curved shape (i.e. it is not a straight line power law). Furthermore we chose a characteristic hardening radius which is lower than in \citet{AgazieBHB_2023} which serves to increase the strength of the turnover in this fiducial model. This turnover is unaffected by changes to the \mmb\ normalization, but a strongly evolving scatter straightens the spectrum out, apparently approaching a power law. This shape change in the evolving scatter model is due to the frequency-dependent increase in GWB amplitude which is a result of the preferential increase in number density of the most massive SMBHs. The impact that a high \mmb\ intrinsic scatter has on the spectrum seems to have a similar, but opposite effect to the presence of environmental hardening. This makes disentangling the contributions from each of these population parameters difficult. Since the GWB data do show a turnover, this could mean that either the \mmb\ scatter cannot evolve strongly or environmental effects may be more important for recovering the spectral shape.

These differences, though large, require high confidence data and precise population statistics to differentiate. Separating these effects is most feasible when only one type of evolution is present and the strength of evolution is great. If either one is subtle (low values of $\alpha_z$ and/or $\varepsilon_z$) or if both types of evolution are present, the problem becomes difficult to disentangle. We compare between models which allow for multiple evolutionary paths of \mmb\ and place constraints using both EM and GWB data in Section \ref{sec:results_em_comp} and Figure \ref{fig:mmb_obs_comp}.

\section{Results}\label{sec:results}

\begin{figure*}[ht]\centering
    \includegraphics[width=\textwidth, keepaspectratio]{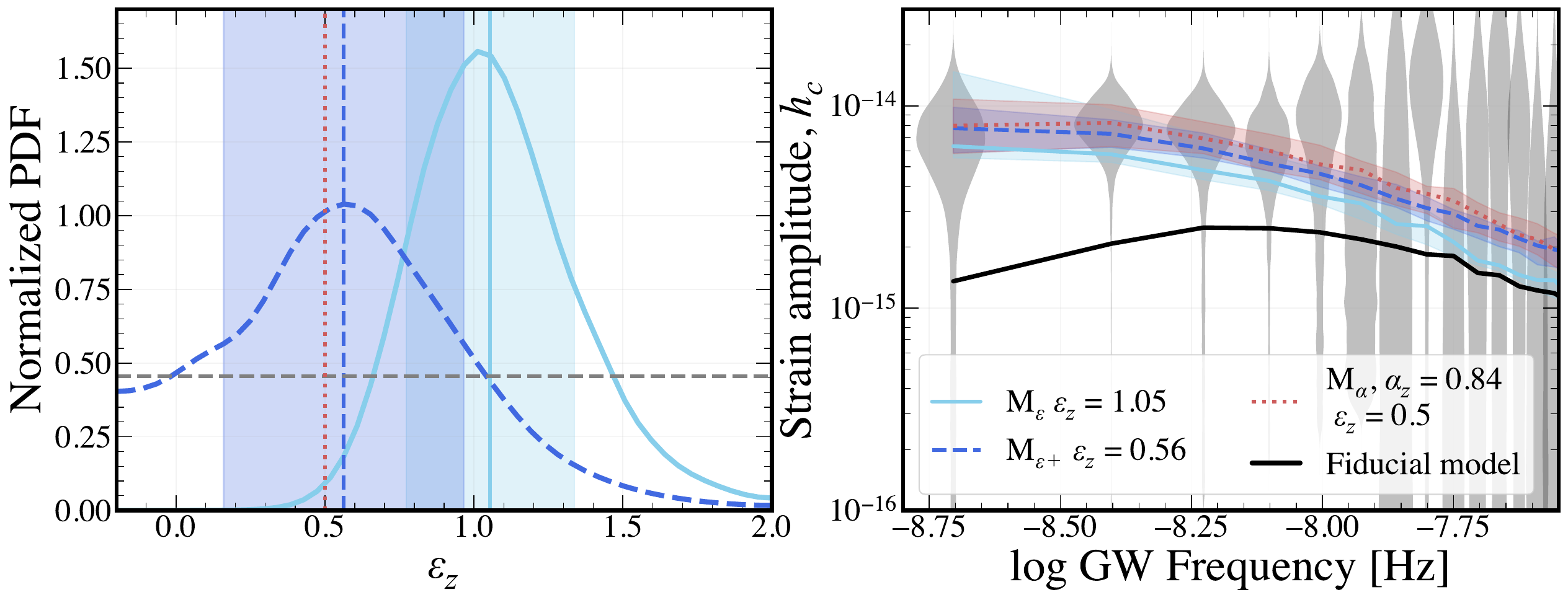}
    \caption{\textit{Left:} The posterior distributions for $\varepsilon_z$ from \mone\ (light blue, solid) and \mtwo\ (dark blue, dashed). The vertical lines represent the median of the posteriors, for \mthree\ (red, dotted) $\varepsilon_z$ was fixed to a value of 0.5 so there is no distribution associated with the median line. The median value of \mone\ is higher than that of \mtwo\ because it had no additional free parameters to increase the amplitude and so $\varepsilon_z$ had to compensate more. \textit{Right:} \rev{The best-fit GWB spectra associated with our three models (blues and red) and the fiducial model (black).} \rev{We see that all \rev{best-fit} models have similar amplitudes and shapes. Both \mtwo\ and \mthree\ lie within the 68\% confidence regions of the data except for the fifth frequency bin whereas \mone\ lies just below the 68\% confidence region of all but the lowest frequency bin. All best-fit models are within the 95\% confidence region for the five frequency bins we fit to.} The low-frequency turnover is present in all three \rev{best-fit} models though it is not strong. Notably, \mone\ lies below the other two models and the bulk of the data only truly aligning with the lowest frequency bin. Interestingly, the spectra of \mtwo\ and \mthree\ are quite similar despite \mtwo\ having no evolution in the \mmb\ normalization.}
    \label{fig:epz_posteriors}
\end{figure*}

\rev{We show the results of all our models in Figure \ref{fig:epz_posteriors}. The black line with the strong turnover is our fiducial model. This model was not sampled, this is simply the GWB spectrum that we would expect when using the default values for every parameter and no evolution in the \mmb\ relation (i.e. the same as the $\varepsilon_z = 0$ and $\alpha_z = 0$ spectra in the right-hand column of Figure \ref{fig:epsilon_z_effects}.} We do not perform fits to the data with a non-evolving \mmb, we instead point readers to \citet{Matt_2026} who used the same fiducial model and did perform non-evolving \mmb\ fits. \rev{To compare the different models' ability to reproduce the GWB data, we perform likelihood ratio tests using the likelihoods we calculate during the fitting process for each spectrum. These are presented at the end of Section \ref{sec:results_alpha_z}.}

\subsection{\Mone: The Isolated Effect of $\varepsilon_z$}\label{sec:results_mone}

Our first model, \mone, had $\varepsilon_z$ as the only free parameter. Every other variable in the models was fixed to its fiducial value. This model isolates the effects of $\varepsilon_z$ on the GWB spectrum. In the top row of Figure \ref{fig:epsilon_z_effects}, we kept all parameters fixed while varying only $\varepsilon_z$. This is the exact setup for \mone\ except with a significantly decreased number of samples \rev{of $\varepsilon_z$}.

The posterior distribution of $\varepsilon_z$ in \mone\ (light blue, solid) is apparently Gaussian and centered on $\varepsilon_z = 1.05 \pm 0.28$, indicating a strong positive evolution of \mmb\ scatter. The amplitude of the spectrum falls just short of the \rev{68\% confidence region of the observed GWB except in the lowest frequency bin}. With only one parameter to vary, this result makes sense; the fiducial spectrum is low in amplitude \rev{and the value of $\varepsilon_z$ correlates positively with GWB amplitude (see Figure \ref{fig:gwb_vs_ez}) therefore positive evolution of \mmb\ scatter is favored}. Since $\varepsilon_z$ has less of an effect on higher frequencies, the ability for this single parameter model to match all five of the lowest frequency bins is limited and the spectrum will systematically under predict the data with increasing discrepancy as frequency increases (assuming agreement in the lowest frequency bin).

\rev{This spectrum is the worst fit to the data out of our three models (see Section \ref{sec:results_alpha_z}).} It is, however, significantly improved over the fiducial model and so, while evolving scatter is not a complete solution, it appears to bridge the majority of the gap between the simplest models and observation. In the following sections we describe how other parameters can resolve the remaining discrepancy.

\subsection{\Mtwo: Effect of $\varepsilon_z$ in the Presence of Additional Degrees of Freedom}\label{sec:results_mtwo}

\Mtwo\ sampled nine total parameters: $\varepsilon_z$, three hardening parameters ($\tau_f$, $\nu_\mathrm{inner}$, and $R_\mathrm{char}$), the local normalization and slope of the \mmb\ relation ($\alpha_0$ and $\beta_0$), and the three local galaxy stellar mass (GSMF) function parameters ($\phi_{*, 0, 0}$, $\phi_{*, 1, 0}$, and $M_{\mathrm{c}, 0}$). For the \mmb\ relation and GSMF parameters we use informative priors based on EM observations \citep{Kormendy_Ho_2013, Leja_2020}, we used uniform priors for the hardening parameters \citep{Kelley_2017, AgazieBHB_2023}.

In Figure \ref{fig:epz_posteriors}, the posterior distribution of $\varepsilon_z$ for \mtwo\ (dark blue, dashed) favors lower values than with \mone, but \rev{92\% of the distribution corresponds to positive values of $\varepsilon_z$ with a median of $\varepsilon_z = 0.56 \pm 0.40$.} \rev{Despite the more modest scatter evolution, the spectrum associated with this model is higher in amplitude than that of \mone\ and is in better agreement with the data, sitting within the 68\% confidence region for the four lowest frequency bins. All GSMF and local \mmb\ posteriors are in good agreement with the priors. \citet{Matt_2026} found that, for a non-evolving \mmb\ relation, posterior distributions for these parameters will deviate significantly from the priors. Since these parameters are well-constrained with EM observations \citep{Leja_2020, Kormendy_Ho_2013}, congruency between posteriors and priors here indicates consistency with EM-based observations. With a weaker \mmb\ scatter evolution (versus \mone) we would expect the amplitude of the GWB to be lower. However the correlation between $\varepsilon_z$ and GWB amplitude is less direct with so many additional free parameters.}

\begin{figure}[ht]\centering
    \includegraphics[width=\columnwidth, keepaspectratio]{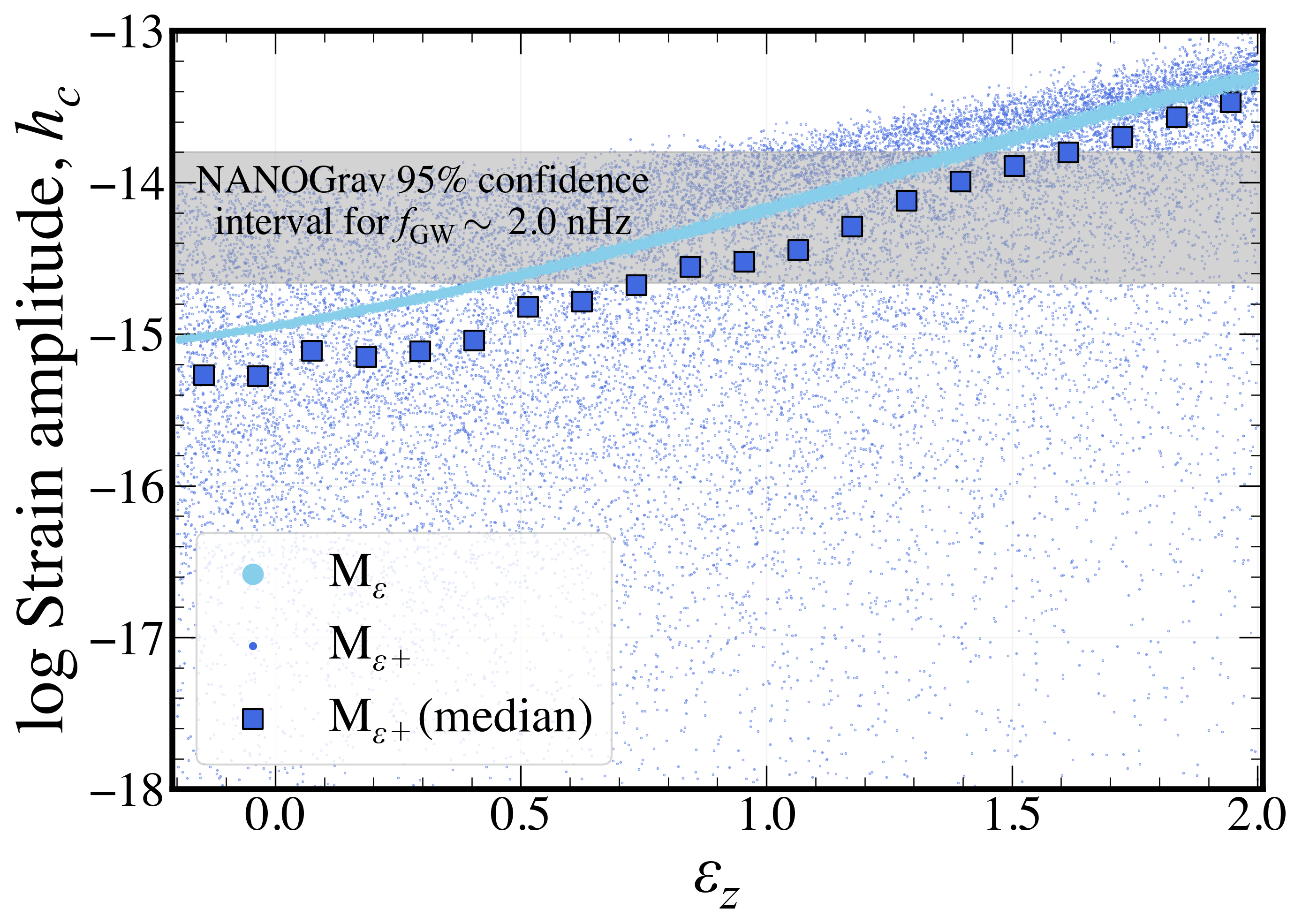}
    \caption{\rev{In this figure we plot the log strain GWB amplitude, $h_c$, measured at the lowest frequency bin $f_\mathrm{gw} \sim 2$ nHz against samples of $\varepsilon_z$ drawn from our prior distribution. We show how the GWB amplitude of a model correlates with the sampled value of $\varepsilon_z$ for \mone\ and \mtwo. The dark blue points with large spread are from \mtwo\ where the blue squares represent \revtwo{the corresponding median log strain amplitude binned} along the x-axis. The light blue line is from \mone, which had $\varepsilon_z$ as the only free parameter. On average, there is a clear positive correlation between the GWB amplitude and the value of $\varepsilon_z$. We allowed many parameters to vary in \mtwo\ and so the correlation between GWB amplitude and $\varepsilon_z$ is less direct. It is still present (blue squares), but there is a larger spread. Other parameters in the model have a strong correlation with the GWB amplitude (e.g. local \mmb\ normalization, $\alpha_0$) so the model does not need to rely solely on $\varepsilon_z$ to reproduce the GWB amplitude. \revtwo{This indirect correlation causes the posterior distribution of $\varepsilon_z$ to spread out, relative to smaller models. Values of $\varepsilon_z$ that did not provide good fits to the GWB data in \mone, can result in a well-fitting spectrum with \mtwo\ because of the effect of other parameters.}}}
    \label{fig:gwb_vs_ez}
\end{figure}

\rev{In Figure \ref{fig:gwb_vs_ez} we show a scatter plot of the GWB amplitude versus the $\varepsilon_z$. As described in \rev{Section} \ref{sec:sam}, for each of the 20,000 samples taken from our prior distributions, we calculate a GWB spectrum. In Figure \ref{fig:gwb_vs_ez} we plot the amplitude of those 20,000 spectra (at the lowest frequency bin, $\log f_\mathrm{GW} \sim -8.7$) for both \mone\ and \mtwo\ versus the value of $\varepsilon_z$ for the associated sample.}

The light blue line corresponds to \mone, which sampled only $\varepsilon_z$, and the dark blue points are from \mtwo\ which sampled nine total parameters. The blue squares represent the median of the dark blue points, binned along $\varepsilon_z$. For \mone, the GWB amplitude is positively correlated with $\varepsilon_z$. Across our prior range the GWB amplitude increases nearly linearly by approximately two orders of magnitude. Since \mone\ only sampled one parameter, there is very little spread in the light blue points (hence they appear as a line). The only difference between any two points with similar values of $\varepsilon_z$ comes from the Poisson noise in \textsc{holodeck}.

\rev{The positive correlation between GWB amplitude and $\varepsilon_z$ is present in \mtwo, however the spread in amplitude values is much larger than in \mone. The average standard deviation of $\log h_c$ is 0.022 for \mone\ and 1.13 for \mtwo.} Because \mtwo\ varied additional parameters to \mone, two points, with near identical values of $\varepsilon_z$, can have a wide range of other parameter values and so the GWB amplitude for those two sample sets can differ. Therefore the dark blue points deviate from the overall trend because of the increased degrees of freedom in the model, but the median trend (blue squares) matches that of \mone\ very closely. Therefore dependence on $\varepsilon_z$ is present in \rev{\mtwo}, but much less \rev{direct} and so there is less reliance on $\varepsilon_z$ to match the GWB compared with \mone. This same behavior affects the posterior distributions for $\varepsilon_z$. The posterior distribution for \mone\ is narrower and takes on higher values on average than that of \mtwo. With multiple free parameters, \mtwo\ can produce a GWB spectrum consistent with the data with different combinations of parameters.

\subsection{\Mthree: Including Multiple Types of \texorpdfstring{\mmb}{black hole mass bulge mass} Evolution}\label{sec:results_alpha_z}

Our main focus in this paper is the \mmb\ intrinsic scatter, however it is possible that both the normalization and scatter of the \mmb\ are changing with time. In Section \ref{sec:impact_of_epz} we saw that changes to normalization and scatter have different impacts on the shape of the GWB spectrum. It is therefore feasible that both types of evolution can provide better fits to the GWB than either one alone. It is also possible that we can differentiate between these models with the available data.

Though their effects are frequency-dependent, both $\alpha_z$ and $\varepsilon_z$ generally scale positively with the GWB amplitude and so there is a \rev{negative correlation between these parameters in our models (correlation coefficient of $-0.85$)}. To mitigate the effect of this degeneracy, we chose to fix the $\varepsilon_z$ parameter to a constant value of 0.5 while allowing $\alpha_z$ to vary. We informed this value of $\varepsilon_z$ based on the posterior distribution of \mtwo. This value of $\varepsilon_z$ is high enough to be a significant evolution without being so high as to dominate the shape or amplitude of the GWB spectrum. This model we refer to as \mthree.

With this model, the posterior distribution for $\alpha_z$ (Figure \ref{fig:alpha_z_posterior}) is 99.9\% positive with a median value of $\alpha_z = 0.84 \pm 0.35$. The associated best-fit spectrum is shown in Figure \ref{fig:epz_posteriors} (dark red, dotted). This spectrum is notably higher in amplitude than our $\varepsilon_z$-only model, but is only minimally different from \mtwo. The \mthree\ best-fit spectrum has a marginally stronger turnover, but overall the single $\alpha_z$ parameter has a similar overall effect to the combined impact of the 8 additional free parameters in \mtwo.

\revtwo{To compare the model fits, we calculate the linear Bayes factors between our models by marginalizing the likelihoods (calculated during fitting) over the prior distributions. In this case, \mthree\ is favored over both other models, though the preference is weak. The Bayes factor in favor of \mthree\ over \mone, calculated across the first five (eleven) frequency bins, is 2.17 (3.77) and the Bayes factor in favor of \mthree\ over \mtwo\ is 1.32 (3.02). The higher frequencies are where the effects of scatter and normalization evolution are most significantly different (e.g., Figure \ref{fig:epsilon_z_effects}) therefore it makes sense that the Bayes factor between \mtwo\ and \mthree\ would change when including these higher frequencies. With more frequencies, the inclusion of an evolving \mmb\ normalization becomes more important for reproducing the observed spectrum. The GWB data, however, are not well constrained beyond the first five frequency bins so we cannot make strong conclusions based on these numbers. In the following \rev{Section} we further evaluate these models in the context of additional EM data.}

\section{Comparisons to \texorpdfstring{\mmb}{black hole mass bulge mass} Relations measured Across Cosmic Time}\label{sec:results_em_comp}

\begin{figure*}[ht]\centering
    \includegraphics[width=\textwidth, keepaspectratio]{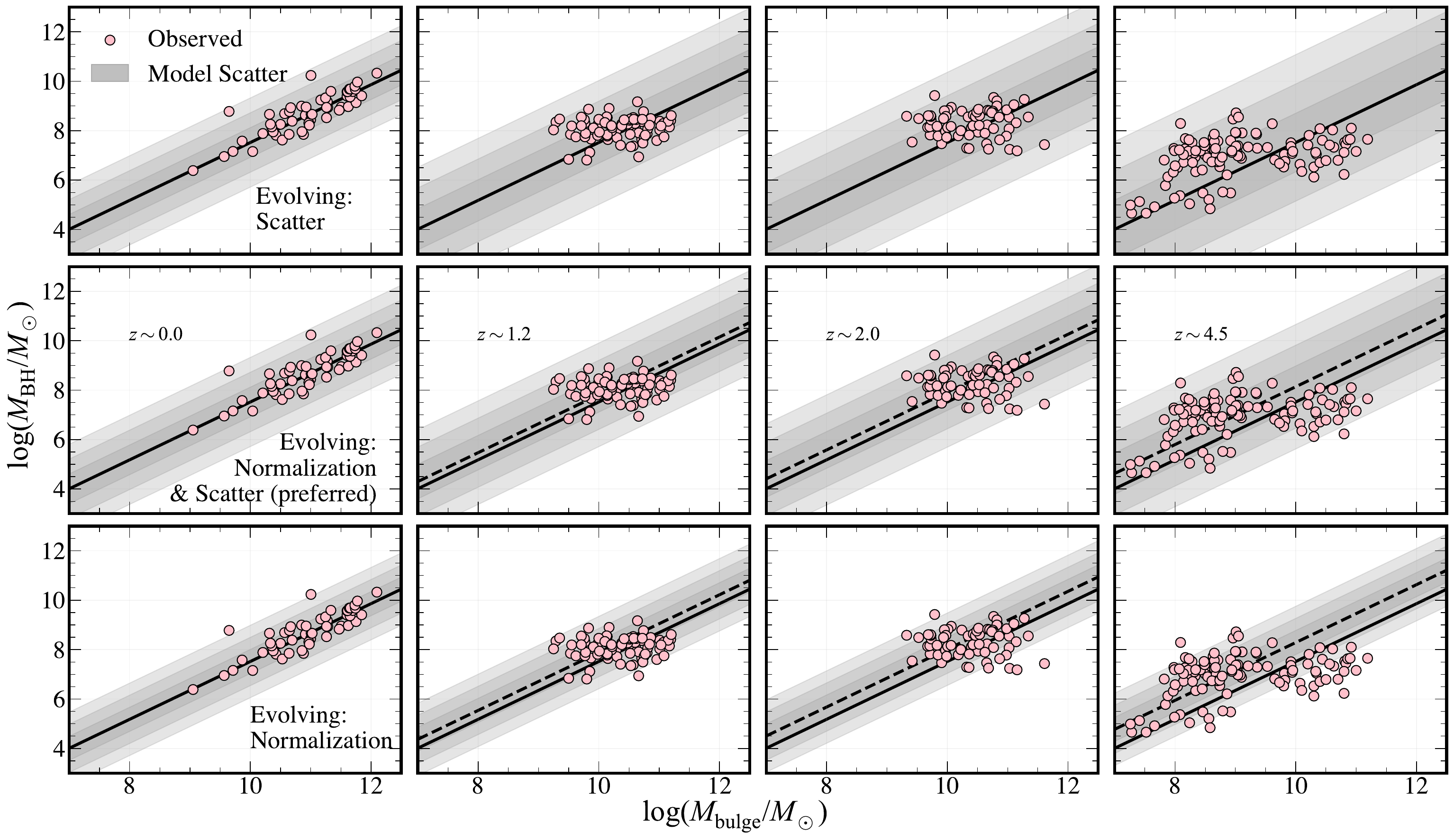}
    \caption{A comparison between our result predictions and observed SMBH and galaxy masses. In gray, we show the 1-, 2-, and 3-$\sigma$ regions of scatter around the \mmb\ relation from \citet{Kormendy_Ho_2013} with three different options for evolution. We add 0.4 dex in quadrature to the intrinsic scatter to approximate measurement error. In all panels the solid black line represents the local \mmb\ relation and the dashed line is the location of the evolving relation. The pink circles in each panel represent EM-observed SMBH--galaxy measurements from the literature. From left to right we include data from \citet{Kormendy_Ho_2013}, \citet{Ding_2020}, \citet{Zhang_2023}, and the final panel includes data from \citet{Sun_2025_corrected}, \citet{Li_2025_normal}, \citet{Jones_2025}, and \citet{Brooks_2025}. \textit{Top row:} The intrinsic scatter of the relation evolves according to the median posterior value from \rev{\mone, $\varepsilon_z = 1.05$}. The normalization is unchanging. \textit{Middle row:} Both the intrinsic scatter and normalization are evolving according to the results from \mthree, $\varepsilon_z = 0.5$, $\alpha_z = 0.84$. \textit{Bottom row:} Only the normalization is evolving here. This represents the best-fit model from \citet{Matt_2026}\rev{, $\alpha_z = 1.04$}. All three cases predict a population of overmassive SMBHs consistent with high-redshift JWST observations without violating local constraints. The two options with evolving scatter also predict an undermassive population, which can reproduce the ``normal mass'' SMBHs found by \citet{Li_2025_normal}. This lower-mass population is not predicted by the high normalization, low scatter model.}
    \label{fig:mmb_obs_comp}
\end{figure*}

\rev{In previous sections, we showed that the GWB data alone may be able to break the degeneracy between types of \mmb\ evolution with improvements to the constraints for high frequency data. Here we investigate how EM data may be able to provide independent constraints.} EM-based studies of the \mmb\ relation outside the local universe exist across a wide range of redshifts. The sample size of observed SMBH–galaxy pairs decreases with increasing redshift due to observational limitations. Now, several years after the launch of JWST, the field has amassed a decent sample of extremely high redshift studies of the \mmb\ relation. \rev{The picture of the $z > 4$ SMBH population is becoming increasingly complex as some studies find SMBH--galaxy pairs that lie significantly above the local \mmb\ relation \citep[e.g.,][]{Pacucci_2023, Yue_2024} while others find pairs that are consistent with (or even below) local measurements \citep[e.g,][]{Ding_2023, Sun_2025_ne, Li_2025_normal}.} This diversity of mass ratios has intriguing implications for the model of SMBH–galaxy coevolution and suggests a wide variety of growth pathways. It additionally hints at possible observational biases which can lead to misinformed conclusions about the entire population. Whether these high-redshift, massive SMBHs represent an overmassive population or the extreme end of a relation, they are not predicted by extrapolations of the local relation. A full analysis of redshift evolution of intrinsic scatter in the \mmb\ relation using only EM data is outside the scope of this paper. Even a simple analysis, however, yields valuable insights. Here we present a qualitative but robust comparison of the predictions from our results and various EM-based SMBH–galaxy mass measurements.

In Figure \ref{fig:mmb_obs_comp} we show the predicted \mmb\ relation with intrinsic scatter that evolves according to our best-fit parameters. \rev{We are interested in distinguishing between these two evolutionary options. Previously, we demonstrated that different types of \mmb\ evolution have different impacts on the GWB spectrum which may someday be differentiable with GWB data alone. We now consider how EM-based measurements of SMBH and galaxy masses may be able to place independent and \rev{complementary} constraints on this problem. For this purpose, we model the evolution of the intrinsic \mmb\ relation in three cases:} (i) a relation with strong scatter evolution, but no normalization evolution (top row, \mone); (ii) a relation with moderate evolution in both scatter and normalization (middle row, \mthree); and (iii) a relation with strong normalization evolution, but no scatter evolution \citep{Matt_2026} (bottom row). The gray regions in each panel are 1-, 2-, and 3-$\sigma$ regions of the intrinsic scatter at that redshift (which is modeled by a normal distribution). \rev{Measurement errors on SMBH masses inferred from luminosity scaling relations cause the apparent \mmb\ scatter to be greater than the intrinsic scatter \citep[typically ranging form 0.2--0.6 dex depending on the relation, e.g.,][]{Vestergaard_2006, Dalla_Bonta_2025}. We therefore added 0.4 dex in quadrature to the modeled intrinsic scatter to account for this. We additionally add the average stellar mass error (0.3 dex multiplied by the \mmb\ slope) in quadrature. These errors are the average reported values across the studies we consider here.}

Overlaid are a selection of observed EM-based SMBH--galaxy pairs with mass measurements from the literature. For all but the highest redshift panel, we only compare our simulated population to one individual study per redshift bin. \rev{The data from \citet{Ding_2020} and \citet{Zhang_2023} are each from a single survey so} we can ignore cross-study systematics which may \rev{otherwise} artificially inflate the scatter of the relation in these bins. There are limited samples for $z > 4$ so we compile several datasets for our highest redshift bin. We carefully selected data from studies which combined several high-redshift SMBH–galaxy mass estimates and corrected for study systematics \citet{Sun_2025_corrected, Li_2025_normal}. We additionally include data from \citet{Jones_2025} and \citet{Brooks_2025}. Except for \citet{Brooks_2025}, these studies all use similar methods for measuring SMBH and galaxy masses, e.g., a delayed-$\tau$ star formation history model, Chabrier initial mass function \citep{Chabrier_2003}, and H$-\alpha$ broad line derived SMBH masses. We include the data from \citet{Brooks_2025} to increase the completeness of our sample, but we recognize this may contaminate our scatter estimates for this bin..

In Figure \ref{fig:mmb_obs_comp} the two models with scatter evolution are more consistent with the EM data than the evolving normalization only model. This is especially apparent at higher redshifts; in the $z \sim 5.0$ panel the model without increasing scatter fails to capture the normal and lower mass SMBH–galaxy pairs. While \mone\ previously was a relatively poor fit to the GWB spectrum, here it provides the best agreement with the EM datasets at all redshifts. The simulated population based on \mtwo\ predicts more overmassive and fewer undermassive SMBHs than we see in the data for $ z \gtrsim 2.0$ though there are at least some SMBHs in every region where EM data are. The model without scatter evolution predicts nearly no SMBH--galaxy pairs below the local relation at $z \sim 5.0$. This model provides excellent fits to the GWB spectrum and, when only considering gravitational wave data, is the best model \citep{Matt_2026}. However, given that pairs have been electromagnetically observed to lie on and below the local relation in this redshift range, we conclude that this model is a poor description of all the available data.

\begin{figure}[ht]\centering
    \includegraphics[width=\columnwidth, keepaspectratio]{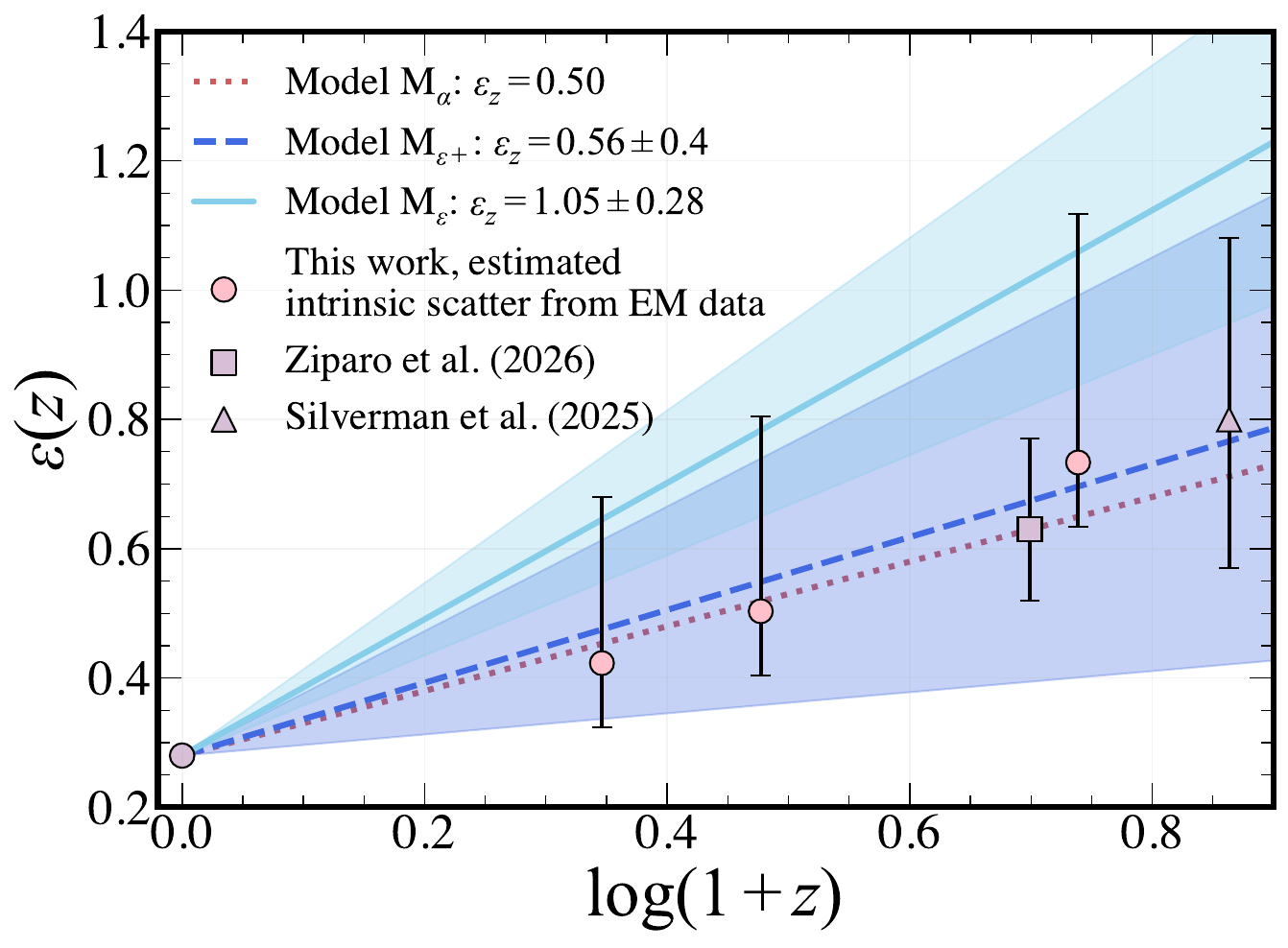}
    \caption{\rev{The value of scatter in the \mmb\ relation versus redshift for the data in Figure \ref{fig:mmb_obs_comp}. The lines represent the predicted evolution of $\varepsilon(z)$ using the median of the posterior for \rev{$\varepsilon_z$ in \mone\ (solid blue), \mtwo\ (dashed blue) and the fixed value in \mthree\ (dotted red) with their 68\% confidence intervals indicated by the shaded regions and in the caption}. We estimate the intrinsic scatter of the EM data in our three non-local redshift bins by calculating the standard deviation of the vertical distance from the local \citet{Kormendy_Ho_2013} relation and subtracting the measurement error in quadrature (pink points). The lowest redshift purple circle is the 0.28 dex value reported by \citet{Kormendy_Ho_2013}. We additionally show two high-redshift scatter estimates from the literature as the purple square \citep{Ziparo_2026} and purple triangle \citep{Silverman_2025}. The estimated and literature scatter values lie close to the model predictions from \mtwo\ and \mthree\ indicating that a moderate scatter evolution is favored by the EM data over the strong evolution in \mone.}}
    \label{fig:ez_vs_z}
\end{figure}

\rev{In Figure \ref{fig:ez_vs_z} we show the predicted value of $\varepsilon(z)$ across redshift from each of our three models. We compare these to estimates of the intrinsic scatter in the three highest redshift bins in Figure \ref{fig:mmb_obs_comp} (pink circles). Our model necessarily matches the \citet{Kormendy_Ho_2013} value (0.28 dex) at $z = 0$ which we show as the left-most purple circle. We additionally show two measurements of the \mmb\ scatter from the literature as the purple square \citep{Ziparo_2026} and purple triangle \citep{Silverman_2025}. We note that the data in \citet{Ziparo_2026} are reported to lie at $ z \gtrsim 4$ and we place it at $z = 4$ in this plot. We measure the total scatter\revtwo{, $\sigma_\mathrm{tot}$,} of the EM data as the standard deviation of the difference between the reported SMBH mass and the local \mmb\ relation. To estimate intrinsic scatter, we subtract the \revtwo{SMBH mass and galaxy stellar/bulge mass measurement error from the total scatter in quadrature. Specifically, we calculate $\sigma_\mathrm{intrinsic}^2 = \sigma_\mathrm{tot}^2 - M_\mathrm{BH, err}^2 - (M_\mathrm{stellar, err} \beta_0)^2$ where  $\beta_0$ is the local slope of the \mmb\ relation.} For each study we take the average reported error if errors are different for each object, but consider the full range in our uncertainties. We do not have bulge masses for the two highest redshift bins. \revtwo{We tested the effect of bulge fraction uncertainty by measuring the scatter of a simulated sample of galaxies with bulge fractions drawn from a uniform distribution. With a lower limit on bulge fraction of 15\%, we determined the typical effect is to inflate the inferred scatter by $\sim 0.1$ dex so we lower the $\sigma_\mathrm{intrinsic}$ estimate by 0.1 dex in both bins}. To account for unknown systematics and the large uncertainties in the stellar mass estimates for \citet{Jones_2025} we additionally subtract off 0.15 dex from \revtwo{$\sigma_\mathrm{intrinsic}$} in the highest redshift bin. \revtwo{This value is motivated by the total spread in the individually measured scatters for each study. We found the difference between the highest and lowest individual scatter to be $\sim 0.3$ dex so we subtracted off 0.15 dex to accommodate this uncertainty.} \revthree{This large range in scatter estimates is impacted by the fact that the sample from \citet{Jones_2025} is made up of 43\% little red dots. The stellar and black hole mass estimates for these objects are highly uncertain so, while we include this data for completeness, we acknowledge that it increases the uncertainty in our scatter measurement in the highest redshift bin. None of the other studies we consider include little red dots in their sample.} The upper limit of our error bars represents the standard deviation of the data before taking any errors into account. The lower limits come from our scatter estimates in the most conservative case for what we assume for errors on mass measurements and bulge fraction uncertainties. We see that the EM-based trend is most similar to the models with more moderate evolution, \mtwo\ and \mthree.}

This semi-qualitative comparison demonstrates the power of a multimessenger analysis. We show that an \mmb\ relation with evolving intrinsic scatter provides better fits to EM data than models with no scatter evolution. When fitting the GWB data, we find that models without \mmb\ normalization fail to recover both the amplitude and shape of the spectrum. Models with evolving normalization only can reproduce the GWB data, but fail to match EM observations. Including both normalization and scatter evolution provide decent fits to the GWB data and make predictions compatible with EM observations. We therefore conclude that an \mmb\ relation which evolves moderately in both normalization and scatter provides the best single model of all available data.

It has been known and discussed for many years that magnitude-limited surveys can bias analyses of the \mmb\ relation toward higher normalizations and shallower slopes \citep{Lauer_2007}. In other words, a population of SMBHs and galaxies that follows a high-scatter \mmb\ relation can appear to follow a higher-normalization \mmb\ relation when sampled with a magnitude-limited survey. Our results suggest that the amplitude of the normalization offset observed at $z > 4$ may not be as large as reported in some of the earlier JWST results \citep{Pacucci_2023}; however, the GWB data do support a non-negligible positive normalization offset. It remains challenging to disentangle systematic and observational biases from the EM data. As more surveys begin probing the lower SMBH mass and lower SMBH–galaxy mass ratio regimes, a dedicated electromagnetic-based study of the intrinsic scatter at high redshift is necessary \citep{Silverman_2025, Ziparo_2026}.

\section{Black Hole Number Density}\label{sec:bhmfs}

\begin{figure*}[ht]\centering
    \includegraphics[width=\textwidth, keepaspectratio]{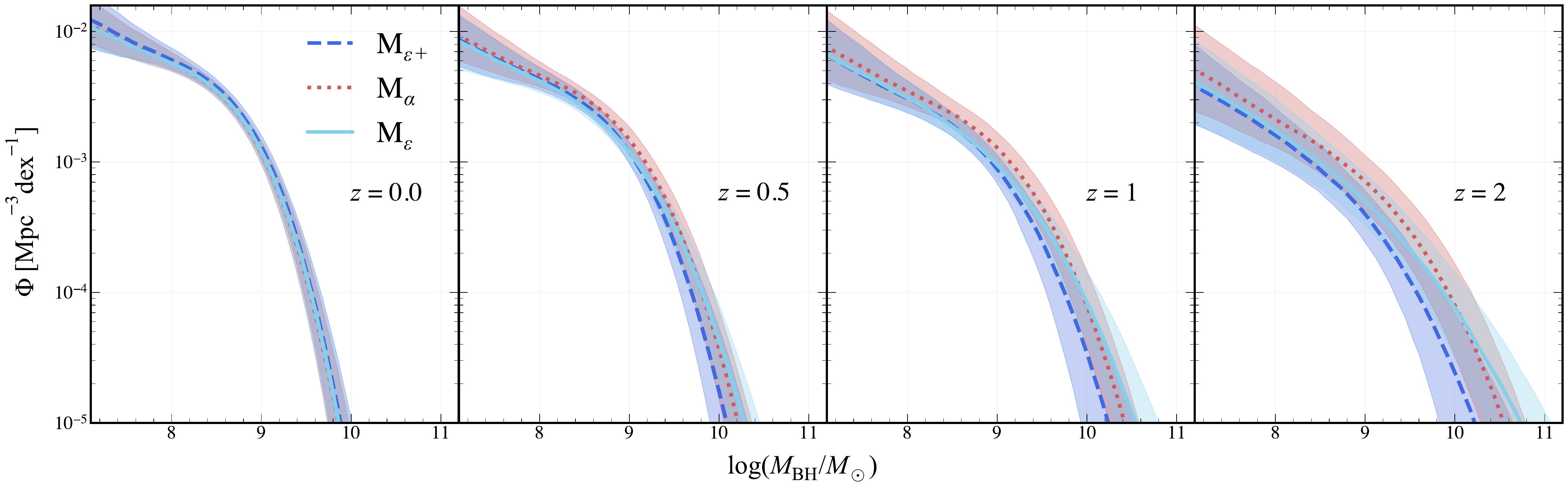}
    \caption{The BHMFs associated with the GWB spectrum best-fit parameters for our models. The solid light blue BHMF represents our model which sampled only the scatter evolution parameter, \mone. The darker blue dashed line represents \mtwo\ and the red dotted line is \mthree. Locally, all three models with evolving scatter are identical by design. Both BHMFs from \mone\ and \mthree\ track each other nearly identically at each redshift with \mthree\ having very slightly higher number densities of $M_\mathrm{BH} \lesssim 10^{10} M_\odot$.  Especially at the highest redshifts, the BHMF from \mtwo\ is generally below all others. This model has a higher GWB amplitude than \mone, but lower SMBH number densities nearly everywhere. The additional parameters in \mtwo, such as the hardening timescales affect the GWB amplitude, but are not reflected in the BHMF. The value of $\varepsilon_z$ is similar between \mtwo\ and \mthree, but the BHMF for \mthree\ is more similar to that of \mone\ because the $\alpha_z$ parameter increases the GWB amplitude via increasing the SMBH number density.}
    \label{fig:bhmfs}
\end{figure*}

The observable nanohertz GWB amplitude is determined from the number of SMBH binaries emitting gravitational waves in the PTA regime. From our fits, we can calculate the implied total SMBH mass distribution which describes the GWB. In Figure \ref{fig:bhmfs} we show the BHMFs associated with each of our best-fit models. Necessarily, all three BHMFs from our evolving models are equivalent at $z = 0$. We \rev{do not} see noticeable differences until $z > 1$ where the BHMFs from \mone\ and \mthree\ have similar number densities to each other, but both higher than the \mtwo\ BHMF. Though the difference appears only slight, the increased number density for $M_\mathrm{BH} \lesssim 10^{10} M_\odot$ seen in \mthree\ leads to the higher GWB spectrum amplitude in \mthree\ versus \mone\ in Figure \ref{fig:epz_posteriors}. \rev{This effect is the same as demonstrated in Figure \ref{fig:epsilon_z_effects} and discussed in Section \ref{sec:impact_of_epz}.}

\begin{figure*}
    \includegraphics[width=\textwidth, keepaspectratio]{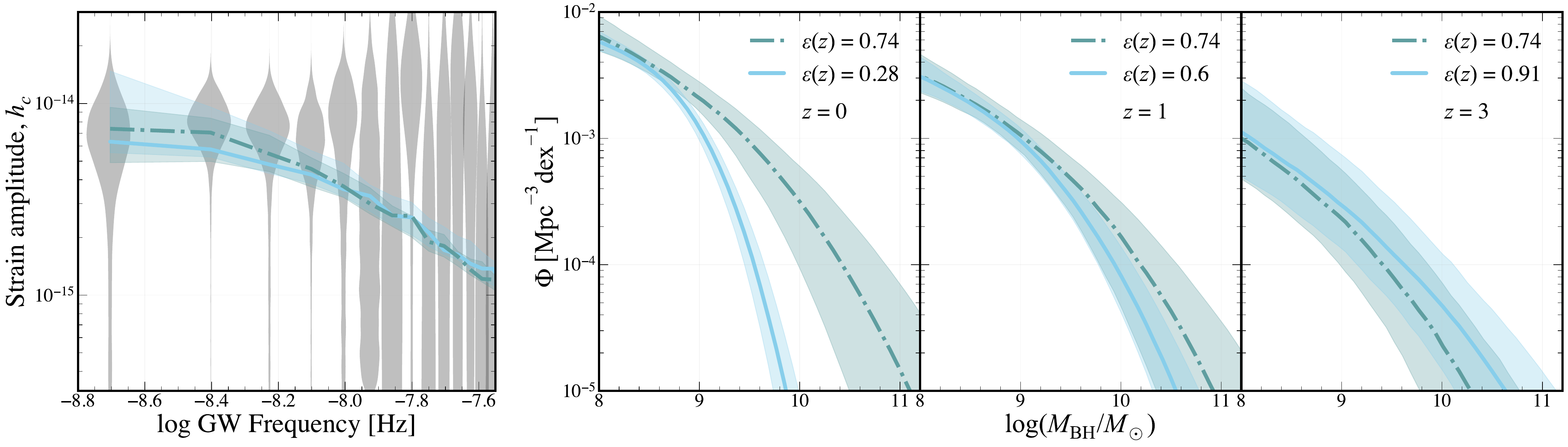}
    \caption{A comparison between the best-fit spectra and associated BHMFs from \mone\ (solid) and a model with non-evolving, but increased scatter (dot-dashed). The BHMF from \mone\ is the same as in Figure \ref{fig:bhmfs}. In the upper right, we show the value of the intrinsic scatter for each model at the relevant redshift, only the value associated with \mone\ changes since the other is non-evolving.}
    \label{fig:eps0_vs_epsz}
\end{figure*}

We consider an additional test case of a model which does not allow for scatter evolution and instead only sampled the local scatter value. It is important to note that this test case is ruled out by the constraints on the local intrinsic scatter, but it is useful to consider the differences between evolving and non-evolving scatter. We compare this model with \mone\ in Figure \ref{fig:eps0_vs_epsz}. Increasing the local scatter matches the GWB just as well as evolving scatter. At $z = 0$ the BHMFs are very different with the locally increased scatter BHMF having a number density over $10\times$ higher than \mone\ for $M_\mathrm{BH} > 10^{10} M_\odot$. This difference is less significant for higher redshifts where the BHMFs are roughly consistent for $z >1$.

\section{Discussion}\label{sec:discussion}

An evolving relationship between SMBH and galaxy bulge mass offers a promising explanation which can unify electromagnetic observations at high and low redshift with gravitational wave data. In this work we focus on the impact of an evolving intrinsic scatter in the \mmb\ relation on predicted SMBH populations and the predicted GWB. Our results show that a high scatter outside the local universe predicts a population of SMBHs which can produce a GWB amplitude consistent with the observed NANOGrav spectrum while also matching the high-redshift ``overmassive'' SMBH population observed by JWST surveys. Our high-scatter models additionally predict an unobserved population of ``undermassive'' black holes which mirror the ``overmassive'' population.

\rev{A significant number of} SMBH masses measured at $z > 4$ from JWST surveys lie significantly above the local \mmb\ relation \citep[e.g., ][]{Pacucci_2023, Stone_2024, Ubler_2023, Maiolino_2024, Harikane_2023, Yue_2024} even after mass corrections \citep{Li_2025_bias, Sun_2025_corrected}. However several studies have claimed that these objects, while real, are not representative of the underlying population and are instead results of magnitude-limited surveys failing to detect normal and undermassive SMBH--galaxy pairs \citep{Li_2025_bias, Silverman_2025, Ziparo_2026}. Indeed there have been some recent results showing that there may exist a population of SMBHs which are not overmassive relative to their hosts \citep[e.g.,][]{Ding_2023, Li_2025_normal, Brooks_2025, Geris_2026}. Neither the local relation nor a high-normalization relation can explain these results. The local relation will under predict the overmassive SMBH -- galaxy pairs while an evolving normalization cannot reproduce the ``normal'' mass pairs. An evolving intrinsic scatter, however, predicts a large spread in mass ratios which predicts the overmassive, normal mass, and undermassive populations.

An intrinsic scatter which is greater in the past and lower at present day has several physical interpretations. It could imply that the local correlation between SMBH mass and galaxy mass is non-causal, but instead a natural outcome of galaxy and SMBH mergers \citep[e.g.][]{Peng_2007, Jahnke_2011, Hirschmann_2010, Hu_2025}. Alternatively, the intrinsic scatter at high-redshift could carry information about SMBH seeding and feedback interactions during SMBH and galaxy growth \citep{Greene_2020, Zhuang_2023, Terrazas_2025}. A high normalization suggests that SMBHs are born from heavy seeds, but a high scatter would suggest multiple seeding pathways are viable. A scatter which decreases with time could indicate that there are a diversity of growth pathways for SMBH--galaxy pairs. In some cases, SMBH growth may drastically outpace the host, where feedback from SMBH accretion can suppress star formation, delaying stellar mass growth leading to a high $M_\mathrm{BH}/M_\mathrm{bulge}$ ratio early on. Eventually, the galaxy must be able to grow its mass without significant change to the SMBH mass such that the $M_\mathrm{BH}/M_\mathrm{bulge}$ ratio reaches the value we observe today. Conversely, the galaxy may grow first and UV radiation from star formation can prevent gas from reaching the galactic center postponing SMBH accretion and leading to an initially low $M_\mathrm{BH}/M_\mathrm{bulge}$ ratio. A high scatter implies that both of these feedback interactions may be present in different SMBH--pairs rather than one dominating over the other on a population level (which could instead present as a high/low \mmb\ normalization in the past).

It has been suggested that active SMBHs may follow a different host--mass relationship than inactive SMBHs \citep[e.g.][]{Reines_Volonteri_2015, Greene_2020}. Our model does not differentiate between active and quiescent SMBHs. If these populations indeed follow different relations, then, observationally, two (or more) low-scatter relations could blend together to appear as one with higher scatter. Our test which sampled only the local intrinsic scatter provided similarly good fits to the GWB data as the evolving scatter model, but implied a large number of local overmassive SMBHs. With a high intrinsic scatter, the most massive SMBHs need not exist exclusively in massive galaxies and so we would expect to have observed this overmassive population by now with large local surveys.

Due to Eddington bias, the impact of this evolving scatter is greatest for the high-mass end of the BHMF. Therefore this high-scatter \mmb\ relation does not significantly change the number density of SMBHs with masses $M_\mathrm{BH} < 10^{9} M_\odot$. The impact on the GWB amplitude is therefore negligible for frequencies $f_\mathrm{gw} \gtrsim 10$ nHz. From the GWB side, we do not see statistically preference for evolving normalization versus scatter. There is strong evidence to support that evolving normalization is the best fit to the GWB data, but it produces a poor model for the EM data so there is either a large discrepancy here or the GWB data are not yet at the point where we can discern between these options.

This work demonstrates the need for improving the precision of GWB data and for increasing the available EM data. Analyses of the GWB can eventually be used to differentiate between different SMBH population models, but are currently limited by the number of reliable frequency bins. Future gravitational wave data releases with well-constrained high-frequency data will greatly improve our ability to distinguish between the effects of \mmb\ scatter and normalization. Furthermore, it is important to perform dedicated statistical analyses of the available EM data to evaluate the effects of observational bias as well as the impact of assumptions made between papers \citep{Silverman_2025, Ziparo_2026}.

The models we present here indicate that an \mmb\ relation which \rev{evolves} both in scatter and normalization provides better fits than either type of individual evolution. A relation which is large in scatter and minimally offset in normalization would be difficult to confirm observationally. Some studies have made great efforts to remove any possible observational bias from their data to measure an \rev{intrinsic \mmb\ normalization offset though the results of these analyses can yield conflicting results. For example, \citet{Zhang_2023} find a $\sim 0.52$ dex increased $M_\mathrm{BH}/M_\star$ ratio at $z \sim 2$ whereas \citet{Sun_2025_corrected} find their $M_\mathrm{BH}/M_\star$ ratio to be consistent with the local value.}

If the \mmb\ relation is changing with time, in any way, it is imperative that we constrain its functional form. The consequences of misestimating the SMBH population at high redshift extend beyond gravitational wave studies. Other relations, such as the \msigma\ relation, predict different SMBH masses at high-$z$ and may be relatively unchanging with time \rev{\citep[e.g.,][]{Matt_2023, Simon_2023, Huber_2025, Cohn_2025}}. Eventually, the GWB can and should be used to test alternative relationships against each other.

\section{Summary}\label{sec:summary}

We used a semi-analytic model to synthesize populations of SMBH binaries and calculate the resulting GWB. \revtwo{We tested the predictions from different versions of the \mmb\ scaling relation against the NANOGrav 15-year GWB spectrum and electromagnetic observational data and placed constraints on how this relationship may be changing with time.} Our results suggest that the intrinsic scatter of the \mmb\ relation, measured locally to be 0.28 dex, must have been higher in the past. A larger intrinsic scatter at high redshift ($\sim 0.7$ dex by $z \sim 5$) predicts a large population of SMBHs with masses well above the \mmb\ relation. This population produces a GWB higher in amplitude than predicted from a non-evolving \mmb\ relation though this amplitude increase is negligible for the highest frequencies. At this point in time the data are not reliable for frequencies higher than $f \sim 10\,\mathrm{nHz}$ so we cannot determine the validity of the lack of amplitude increase in this regime. A high intrinsic scatter can explain the population of overmassive SMBHs found in various JWST observations, and predicts a population of SMBHs with masses both on and far below the local relation which is not yet observed.

We considered the different impacts between evolving normalization and scatter. We find that both an evolving normalization and scatter provides a better fit to the GWB data than only evolving scatter and is not ruled out by the EM observations. \rev{Considering the population of SMBHs at $z > 4$ we use here \citep{Li_2025_normal, Jones_2025, Brooks_2025, Sun_2025_corrected}}, a low-scatter high-normalization \mmb\ relation fails to predict the underlying SMBH population at high-redshifts despite being a great fit to the GWB data. We therefore conclude that the intrinsic scatter of the \mmb\ relation must be higher in the past to explain high-redshift electromagnetic observations and the normalization must have been higher to explain the gravitational wave data.

\begin{acknowledgments}

C.M. is grateful for support from a Frontera Computational Science Fellowship and computational resources from the Texas Advanced Computing Center at the University of Texas at Austin.
The authors would like to thank Jinyi Yang and Feige Wang for helpful discussions. \rev{The authors also wish to thank the anonymous referee for their thoughtful review and suggestions.}
\input{acks}

Anishinaabeg gaa bi dinokiiwaad temigad manda Michigan Kichi Kinoomaagegamig. Mdaaswi nshwaaswaak shi mdaaswi shi niizhawaaswi gii-sababoonagak, Ojibweg, Odawaag, minwaa Bodwe’aadamiig wiiba gii-miigwenaa’aa maamoonjiniibina Kichi Kinoomaagegamigoong wi pii-gaa aanjibiigaadeg Kichi-Naakonigewinning, debendang manda aki, mampii Niisaajiwan, gewiinwaa niijaansiwaan ji kinoomaagaazinid.  Daapanaming ninda kidwinan, megwaa minwaa gaa bi aankoosejig zhinda akiing minwaa gii-miigwewaad Kichi-Kinoomaagegamigoong aanji-daapinanigaade minwaa mshkowenjigaade.

The University of Michigan is located on the traditional territory of the Anishinaabe people. In 1817, the Ojibwe, Odawa, and Bodewadami Nations made the largest single land transfer to the University of Michigan.  This was offered ceremonially as a gift through the Treaty at the Foot of the Rapids so that their children could be educated. Through these words of acknowledgment, their contemporary and ancestral ties to the land and their contributions to the University are renewed and reaffirmed.

\end{acknowledgments}

\begin{contribution}


\rev{C.M.\ led the analysis and investigation of this project as well as the writing and editing of this manuscript and produced all figures and models. C.M.\ and K.G.\, were responsible for the conceptualization and supervision of this work. K.G. provided useful feedback and comments on the manuscript and aided in scientific interpretation of the results. Additional NANOGrav members are listed in alphabetical order. Each contributed toward the collaboration-wide endeavor of pulsar timing which culminated in evidence for the GWB. The work presented in this paper benefits from data collected and analyzed in the course of this search and would not be possible without this large-scale coordination.}


\end{contribution}

%



\appendix

\begin{figure*}[ht]\centering
    \includegraphics[width=\textwidth, keepaspectratio]{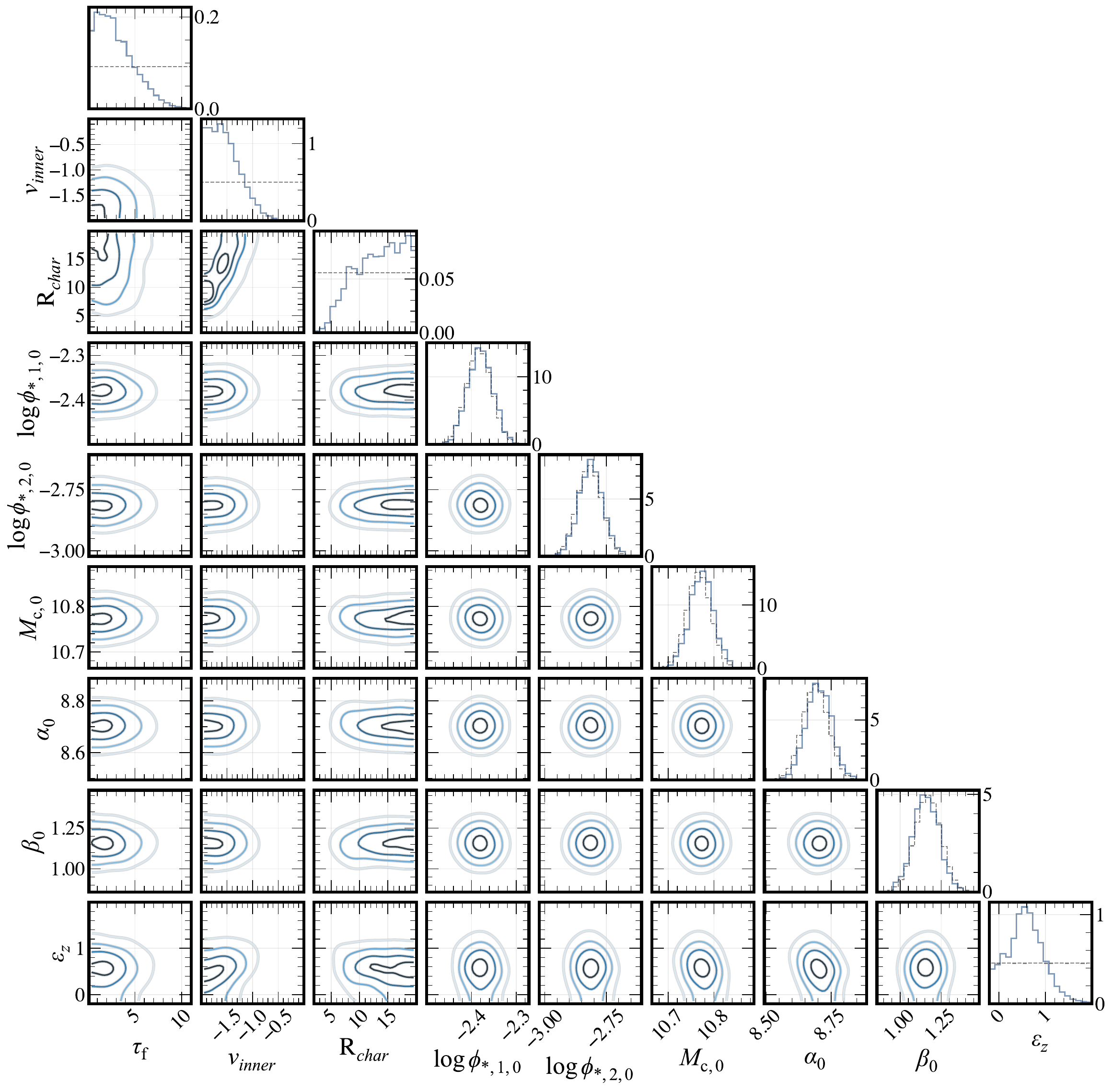}
    \caption{Corner plot of posteriors from \mtwo. The posterior for $\varepsilon_z$ is additionally shown in Figure \ref{fig:epz_posteriors}.}
    \label{fig:asn_corner_scatter}
\end{figure*}

Here we include the posteriors for parameters in \mtwo\ and \mthree\ which were not discussed in the main text. The posterior for $\varepsilon_z$ from \mtwo\ is shown in the lower-right panel of Figure \ref{fig:asn_corner_scatter}; this is the same distribution as seen previously in Figure \ref{fig:epz_posteriors}.

\rev{In Figure \ref{fig:asn_corner_scatter}, we see that the posterior distributions (blue, solid) are nearly identical to the priors (gray, dashed) for all our \rev{Gaussian} priors. These priors are based on electromagnetic observations for the local galaxy stellar mass function \citep[$\log \phi_{*, 1,0}$, $\log \phi_{*, 2,0}$, and $M_{\mathrm{c} ,0}$][]{Leja_2020} and the \mmb\ relation \citep[$\alpha_0$ and $\beta_0$][]{Kormendy_Ho_2013}. The fact that the posteriors recover the priors here indicates that this model is in good agreement with EM observational constraints.}

\rev{All three of the SMBH binary hardening parameters push against their prior bounds. The lower limit on the hardening timescale, $\tau_f$ is 0 Gyr and low values of this parameter were also seen by \citet{AgazieBHB_2023}. $\nu_\mathrm{inner}$ is the slope of the power--law associated with stellar scattering. The posterior returns low values, but the lower bound, $-2$ is sufficiently small to minimize that term and lowering this bound has no significant effect. Finally, $\mathrm{R}_\mathrm{char}$ indicates the point at which the BH binary hardening transitions from dynamical friction into stellar scattering. The physically motivated value is on the order of 10 pc \citep{Kelley_2017}. Our prior range is 2--20 pc and the posterior distribution returns high values within this range. There is a degeneracy between $\mathrm{R}_\mathrm{char}$ and $\nu_\mathrm{inner}$. When we allow $\mathrm{R}_\mathrm{char}$ to go much higher, $\nu_\mathrm{inner}$ also returns higher values. For example, \citet{AgazieBHB_2023} used a fixed value of $\mathrm{R}_\mathrm{char} = 100$ pc and they found $\nu_\mathrm{inner}= -0.5$. The difference between these results is interesting and has been the focus of other studies \citep[e.g.,][]{Harris_2026}, but do not have a significant impact on our conclusions about the redshift evolution of the \mmb\ relation.}

\begin{figure}[ht]\centering
    \includegraphics[width=0.5\columnwidth, keepaspectratio]{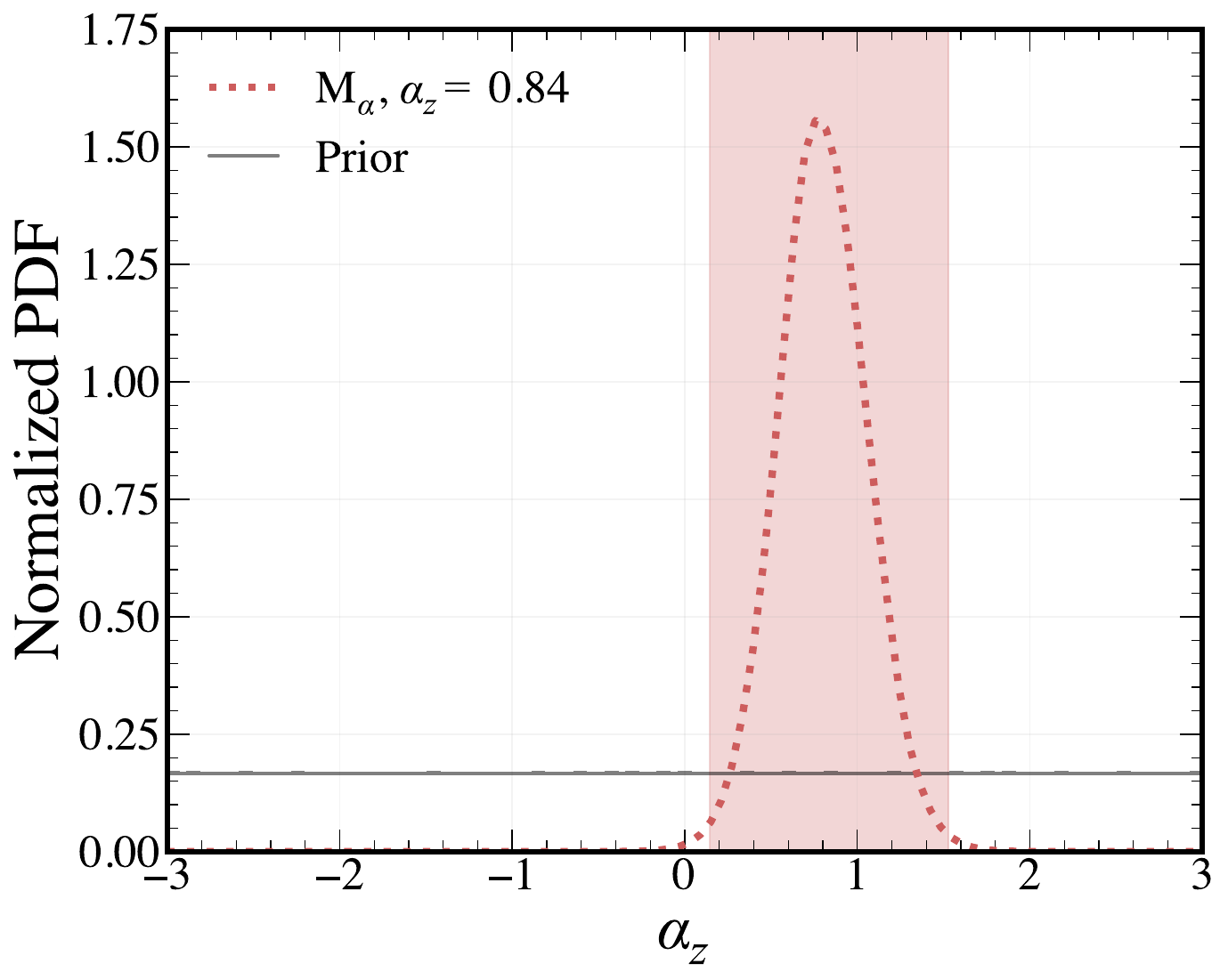}
    \caption{Histogram of the posterior for $\alpha_z$ in \mthree. The red shaded region shows the 68\% confidence interval and the horizontal grey line represents the uniform prior $-3.0 < \alpha_z < 3.0$. This was the only free parameter in this model. All other parameters were fixed to their fiducial values except $\varepsilon_z$ which was fixed to 0.5.}
    \label{fig:alpha_z_posterior}
\end{figure}

The posterior distribution for $\alpha_z$ from \mthree\ is shown in Figure \ref{fig:alpha_z_posterior}. The shape is apparently Gaussian and has a median value of $\alpha_z = 0.84 \pm 0.35$. The distribution is 99.9\% positive indicating a strong preference for positive \mmb\ normalization evolution. This model sampled only $\alpha_z$ and had all other parameters fixed to their fiducial value except $\varepsilon_z = 0.5$. This evolution is a moderately good model for both the GWB and EM datasets we consider.

\bibliography{ref}{}
\bibliographystyle{aasjournalv7}



\end{document}

%% file: authors_scatter_arx.tex
\author{Cayenne Matt \orcid{0000-0002-9710-6527}}
\affiliation{Department of Astronomy and Astrophysics, University of Michigan, Ann Arbor, MI 48109, USA}
\email[show]{cayenne@umich.edu}

\author{Kayhan G\"{u}ltekin \orcid{0000-0002-1146-0198}}
\affiliation{Department of Astronomy and Astrophysics, University of Michigan, Ann Arbor, MI 48109, USA}
\email{kayhan@umich.edu}

\author{Gabriella Agazie \orcid{0000-0001-5134-3925}}
\affiliation{Center for Gravitation, Cosmology and Astrophysics, Department of Physics and Astronomy, University of Wisconsin-Milwaukee,\\ P.O. Box 413, Milwaukee, WI 53201, USA}
\email{gabriella.agazie@nanograv.org}
\author{Nikita Agarwal }
\affiliation{Department of Physics and Astronomy, West Virginia University, P.O. Box 6315, Morgantown, WV 26506, USA}
\affiliation{Center for Gravitational Waves and Cosmology, West Virginia University, Chestnut Ridge Research Building, Morgantown, WV 26505, USA}
\email{nikita.agarwal@nanograv.org}
\author{Akash Anumarlapudi \orcid{0000-0002-8935-9882}}
\affiliation{Department of Physics and Astronomy, University of North Carolina, Chapel Hill, NC 27599, USA}
\email{akasha@unc.edu}
\author{Anne M. Archibald \orcid{0000-0003-0638-3340}}
\affiliation{Newcastle University, NE1 7RU, UK}
\email{anne.archibald@nanograv.org}
\author{Zaven Arzoumanian \orcid{0009-0008-6187-8753}}
\affiliation{X-Ray Astrophysics Laboratory, NASA Goddard Space Flight Center, Code 662, Greenbelt, MD 20771, USA}
\email{zaven.arzoumanian@nanograv.org}
\author{Jeremy G. Baier \orcid{0000-0002-4972-1525}}
\affiliation{Department of Physics, Oregon State University, Corvallis, OR 97331, USA}
\email{jeremy.baier@nanograv.org}
\author{Paul T. Baker \orcid{0000-0003-2745-753X}}
\affiliation{Department of Physics and Astronomy, Widener University, One University Place, Chester, PA 19013, USA}
\email{paul.baker@nanograv.org}
\author{Bence B\'{e}csy \orcid{0000-0003-0909-5563}}
\affiliation{Institute for Gravitational Wave Astronomy and School of Physics and Astronomy, University of Birmingham, Edgbaston, Birmingham B15 2TT, UK}
\email{bence.becsy@nanograv.org}
% ...

%% file: acks.tex
L.B.\ acknowledges support from the National Science Foundation under award AST-2307171 and from the National Aeronautics and Space Administration under award 80NSSC22K0808.
P.R.B.\ is supported by the Science and Technology Facilities Council, grant number ST/W000946/1.
S.B.\ gratefully acknowledges the support of a Sloan Fellowship, and the support of NSF under award \#1815664.
The work of R.B., R.C., X.S., J.T., and D.W.\ is partly supported by the George and Hannah Bolinger Memorial Fund in the College of Science at Oregon State University.
M.C.\ acknowledges support by the European Union (ERC, MMMonsters, 101117624).
Support for this work was provided by the NSF through the Grote Reber Fellowship Program administered by Associated Universities, Inc./National Radio Astronomy Observatory.
H.T.C.\ acknowledges funding from the U.S. Naval Research Laboratory.
Pulsar research at UBC is supported by an NSERC Discovery Grant and by CIFAR.
K.C.\ is supported by a UBC Four Year Fellowship (6456).
M.E.D.\ acknowledges support from the Naval Research Laboratory by NASA under contract S-15633Y.
T.D.\ and M.T.L.\ received support by an NSF Astronomy and Astrophysics Grant (AAG) award number 2009468 during this work.
E.C.F.\ is supported by NASA under award number 80GSFC24M0006.
K.A.G.\ and S.R.T.\ acknowledge support from an NSF CAREER award \#2146016.
D.C.G.\ is supported by NSF Astronomy and Astrophysics Grant (AAG) award \#2406919.
A.D.J.\ acknowledges support from the Caltech and Jet Propulsion Laboratory President's and Director's Research and Development Fund.
A.D.J.\ acknowledges support from the Sloan Foundation.
N.La.\ was supported by the Vanderbilt Initiative in Data Intensive Astrophysics (VIDA) Fellowship.
Part of this research was carried out at the Jet Propulsion Laboratory, California Institute of Technology, under a contract with the National Aeronautics and Space Administration (80NM0018D0004).
D.R.L.\ and M.A.M.\ are supported by NSF \#1458952.
M.A.M.\ is supported by NSF \#2009425.
C.M.F.M.\ was supported in part by the National Science Foundation under Grants No.\ NSF PHY-1748958 and AST-2106552.
A.Mi.\ acknowledges support from a Royal Society University Research Fellowship (URF-R1-251896)
The Dunlap Institute is funded by an endowment established by the David Dunlap family and the University of Toronto.
K.D.O.\ was supported in part by NSF Grant No.\ 2207267.
T.T.P.\ acknowledges support from the Extragalactic Astrophysics Research Group at E\"{o}tv\"{o}s Lor\'{a}nd University, funded by the E\"{o}tv\"{o}s Lor\'{a}nd Research Network (ELKH), which was used during the development of this research.
P.P.\ and S.R.T.\ acknowledge support from NSF AST-2007993.
H.A.R.\ is supported by NSF Partnerships for Research and Education in Physics (PREP) award No.\ 2216793.
S.M.R.\ and I.H.S.\ are CIFAR Fellows.
Portions of this work performed at NRL were supported by ONR 6.1 basic research funding.
J.D.R.\ also acknowledges support from start-up funds from Texas Tech University.
S.C.S.\ and S.J.V.\ are supported by NSF award PHY-2011772.
J.S.\ is supported by an NSF Astronomy and Astrophysics Postdoctoral Fellowship under award AST-2202388, and acknowledges previous support by the NSF under award 1847938.
O.Y.\ is supported by the National Science Foundation Graduate Research Fellowship under Grant No.\ DGE-2139292.